\documentclass[aps,prl,twocolumn,superscriptaddress,showpacs]{revtex4}
\usepackage[latin1]{inputenc}
\usepackage{graphicx}
\usepackage{amsmath,amsfonts}

\begin{document}

\title{Extraordinary optical transmission without plasmons: the \emph{s}-polarization case}

\author{Esteban Moreno}
\email[Electronic address: ]{esteban.moreno@uam.es}
\affiliation{Departamento de Física Teórica de la Materia
Condensada, Universidad Autónoma de Madrid, E-28049 Madrid, Spain}

\author{L. Martín-Moreno}
\affiliation{Departamento de Física de la Materia Condensada,
Universidad de Zaragoza-CSIC, E-50009 Zaragoza, Spain}

\author{F. J. García-Vidal}
\affiliation{Departamento de Física Teórica de la Materia
Condensada, Universidad Autónoma de Madrid, E-28049 Madrid, Spain}

\date{\today}

\begin{abstract}
It is shown that extraordinary optical transmission through
perforated metallic films is possible for \emph{s}-polarization.
Although surface plasmons do not exist for this polarization,
their role can be played by a surface wave sustained by a thin
dielectric layer on top of the metallic film. This confirms that
the existence of a surface wave, whatever its nature, is the
responsible for the extraordinary optical transmission phenomenon.
\end{abstract}

\pacs{78.66.Bz, 42.79.Dj, 71.36.+c, 73.20.Mf}

\maketitle

Already in the original experimental paper~\cite{Ebbesen}, surface
plasmons were pointed to as the reason of extraordinary optical
transmission (EOT) through two-dimensional (2D) arrays of
subwavelength holes in optically thick metallic films. The initial
EOT theoretical
models~\cite{Schroter,Porto,Treacy,Tan,Lalanne,Collin} considered
1D arrays of subwavelength slits. This is a much simpler system
where both polarizations (\emph{s} and \emph{p}) are decoupled.
Moreover, the modes supported by the slits in
\emph{p}-polarization are very different from those supported by
the holes~\cite{Popov}, the former not having cutoff wavelength.
Despite these limitations, the models distinctly showed that EOT
only occurs for \emph{p}-polarization (magnetic field parallel to
the slits), the only one for which surface plasmons are possible.
Later~\cite{Luis}, better suited 2D models also supported the
surface plasmon picture. Shortly after, it was realized that
simulations with perfect metals also displayed EOT and, since flat
perfect conductors do not sustain surface plasmons, this was
unexpected. Such behavior has been recently cleared
up~\cite{Pendry,FJ} by showing that corrugation (with grooves,
holes, or dimples) of perfect metals gives rise to surface
electromagnetic modes with a plasmon-like behavior (so called
spoof plasmons) for the \emph{p}-polarization case. Thus, all
mentioned instances of EOT in metals, and in other systems, e.g.,
photonic crystals~\cite{Moreno1y2}, are mediated by surface waves.

Let us now solely consider the 1D \emph{s}-polarization case. As
said, plasmons in real metals are not possible for this
polarization. An analysis similar to~\cite{FJ} shows that surface
structuring of perfect metals does not produce \emph{s}-polarized
spoof plasmons, the reason being the different kind of boundary
conditions. Since \emph{s}-polarization surface waves are excluded
for metals, it seems that EOT is restricted to
\emph{p}-polarization. In this paper we demonstrate that addition
of a thin dielectric film on the metal interface creates a surface
wave that allows for EOT in the ``wrong'' \emph{s}-polarization
case.

The considered system consists of a thin metallic film embedded in
vacuum, pierced by a periodic slit array with periodicity
$\Lambda$. The film thickness is $t=0.02\Lambda$, and the slit
width is $w=0.22\Lambda$, unless otherwise explicitly stated. The
possibility of supporting a surface wave is achieved by adding a
dielectric layer ($\epsilon_\textrm{r}=4$ is taken for proof of
principle purposes) on the metal film, with thickness $h$. Almost
all results presented here correspond to \emph{s}-polarization
(electric field parallel to the slits). We are interested in the
subwavelength regime, i.e., the slits are narrow and the modes
guided inside the slits are evanescent. Notice that in spite of
the geometric similarity to the case considered in~\cite{Christ},
here we are in a very different regime: the transmittance of our
system without dielectric film is negligible, whereas
in~\cite{Christ} transmittance without film is nearly 1 because
the slits are very wide. For the sake of simplicity our models
consider perfect conductors in most of the paper, which is
undoubtedly appropriate for microwave frequencies. Nevertheless,
in order to show that the mechanism is also valid for the optical
regime, a more realistic dispersive and absorptive dielectric
constant $\epsilon_\textrm{r, metal}(f)$, $f$ standing for the
frequency, is employed towards the end of the paper. All
simulations have been carried out with the multiple multipole
(MMP) method~\cite{Christian}.

\begin{figure}[t]
\includegraphics{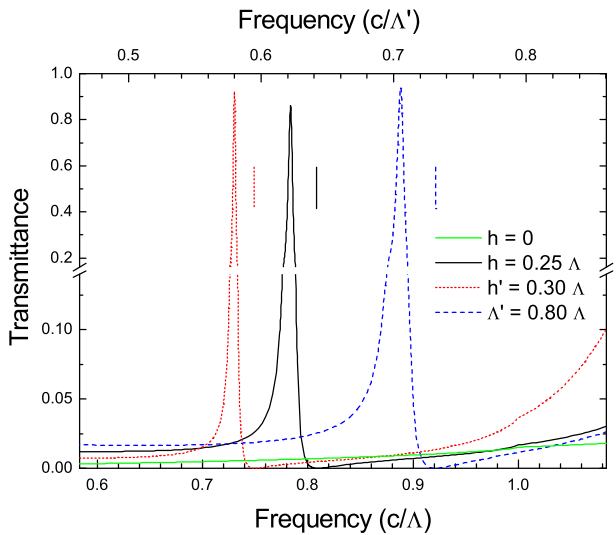}
\caption{\label{EOT}(color online) Transmittance as a function of
frequency for a \emph{s}-polarized plane wave at normal incidence
(the units are $c/\Lambda$, $c$ being the speed of light in
vacuum). Green line: without dielectric film ($h=0$). Black, red,
and blue lines: with dielectric film ($h\neq 0$). The parameters
are described in the main text. The short vertical segments mark
the positions of the frequencies, $f_\textrm{unpert}$, satisfying
Eq.~(\ref{condition}) for the various geometric parameters, see
main text.}
\end{figure}

The green and black lines in Fig.~\ref{EOT} represent the 0-order
diffracted transmittance for orthogonal incidence in the cases
without dielectric layer ($h=0$) or with it ($h=0.25\Lambda$),
respectively. Without dielectric film the transmittance is nearly
zero (less than 2\%), but EOT is observed as soon as a thick
enough film is added (transmittance reaches 86\%). The black line
displays a resonant feature including a maximum and a minimum. The
spectral position of this feature depends on the geometric
parameters and $\epsilon_\textrm{r}$ in the usual way in EOT.
Namely, if we consider the \emph{unperturbed} surface wave guided
by the dielectric film on top of a continuous metallic substrate
(i.e., without slits), we have checked that the resonant feature
is close to the frequency $f_\textrm{unpert}$ that satisfies the
condition
\begin{equation}\label{condition}
\lambda_\textrm{sw}(f_\textrm{unpert}) = \Lambda,
\end{equation}
$\lambda_\textrm{sw}(f)$ being the  modal wavelength of the
unperturbed surface wave. Below we will be more precise about the
relation between the position of the maximum, $f_\textrm{max}$,
minimum, $f_\textrm{min}$, and $f_\textrm{unpert}$. Since the
surface wave dispersion relation, $\lambda_\textrm{sw}(f)$,
depends on the film thickness $h$ and dielectric constant
$\epsilon_\textrm{r}$, the position of the resonant feature
depends on $h$, $\epsilon_\textrm{r}$, and $\Lambda$.
Figure~\ref{EOT} also displays an example varying $h$ and keeping
the other parameters unaltered ($h'=0.30\Lambda$, red line), and
another example varying $\Lambda$ and keeping the other parameters
unaltered ($\Lambda'=0.8\Lambda$; $t=0.025\Lambda'$,
$w=0.275\Lambda'$, $h=0.3125\Lambda'$, blue line).
Equation~(\ref{condition}) gives the location of the resonant
feature in all cases (see the short vertical segments in
Fig.~\ref{EOT}). Notice that, in a sense, the system considered
here is more akin to the original 2D hole array studied
in~\cite{Ebbesen} than the first 1D theoretical models in
\emph{p}-polarization~\cite{Schroter,Porto,Treacy,Lalanne,Collin}.
The reason is that the slits are subwavelength and therefore, for
\emph{s}-polarization, the slit modes are evanescent as it also
happens in the hole array case. In~\cite{Ebbesen} the position of
the resonant frequencies depends on the substrate dielectric
constant, whereas here this role is played by $h$ (and/or
$\epsilon_\textrm{r}$).


If the frequencies $f_\textrm{max}$ and $f_\textrm{min}$ in
Fig.~\ref{EOT} are confronted with the frequency
$f_\textrm{unpert}$, one realizes that $f_\textrm{min}$ is always
very close to $f_\textrm{unpert}$. Such a circumstance sparked a
controversy for the original \emph{p}-polarization EOT models
about the role of the surface plasmon~\cite{Cao}. Since
$f_\textrm{min}\approx f_\textrm{unpert}$ it was argued that the
surface plasmon is only responsible for the minimum in
transmission and therefore it plays a negative role in EOT. In the
original 1D models of EOT, plasmons lay very close to the light
line (for the considered frequency regime), and therefore the
resonance was very close to the Rayleigh anomaly, further
obscuring the subject. In the system considered here the surface
wave can be tailored so that the mode is not close to the light
line and the analysis is easier. This is clear in Fig.~\ref{EOT}
where the resonance is far from the Rayleigh anomaly, occurring
for $f_\textrm{Rayleigh} = c / \Lambda$ (where $c$ is the speed of
light in vacuum).

\begin{figure}[b]
\includegraphics{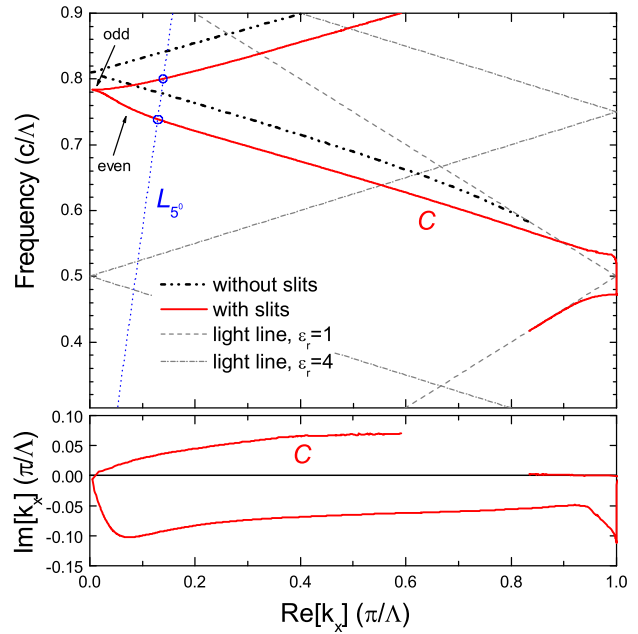}
\caption{\label{bands}(color online) Band structure of the
considered system. Black line: folded dispersion relation of the
unperturbed (i.e., without slits) surface wave. Red line ($C$):
dispersion relation of the actual system with slits. Thin grey
lines: folded light lines in $\epsilon_\textrm{r}=1$ and
$\epsilon_\textrm{r}=4$. The chosen parameters and the blue line
$L_{\theta=5^{\circ}}$ are described in the main text.}
\end{figure}

\begin{figure}[t]
\includegraphics{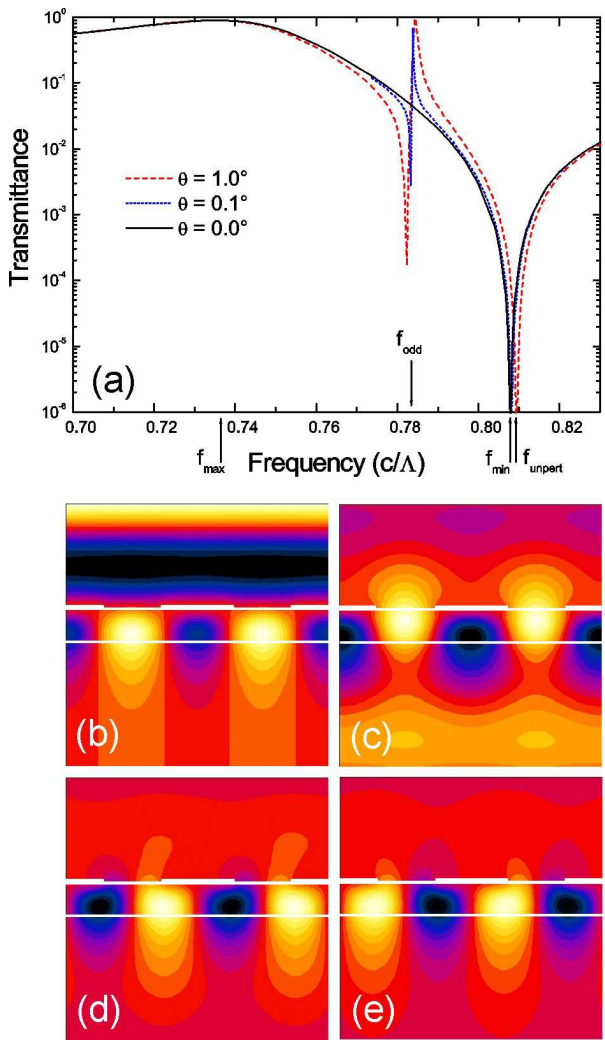}
\caption{\label{oddmode}(color online) (a) Transmittance as a
function of frequency for a \emph{s}-polarized plane wave at
normal and slightly off-normal incidences. Black line: normal
incidence ($\theta=0^{\circ}$). Red line: $\theta=1.0^{\circ}$.
Blue line: $\theta=0.1^{\circ}$. The parameters are described in
the main text. Panels (b)-(c) display the electric field for the
case $\theta=0^{\circ}$ at $f_\textrm{min}$ and $f_\textrm{max}$,
respectively. Panels (d)-(e) display the electric field for the
case $\theta=0.1^{\circ}$ at  $f_\textrm{odd;min}$ and
$f_\textrm{odd;max}$, respectively [the size of the plots is
$2\Lambda \times 2\Lambda$; the color scales are different in the
panels (b)-(e); the plane wave impinges from the top].}
\end{figure}

To shed some light on this issue we have computed the complex band
structure of our system (Fig.~\ref{bands}). In order to better
separate the frequencies of the various modes involved, the
following parameters were chosen: $t=0.02\Lambda$,
$w=0.46\Lambda$, and $h=0.25\Lambda$, i.e., the slit is now wider
than previously, but it remains in the subwavelength regime. The
band structure is obtained by searching for the poles of the
scattering amplitude in the complex plane $k_x$
(see~\cite{Moreno3} for details; similar diagrams are found in
other contexts~\cite{Brundrett}). From Fig.~\ref{bands} it is
possible to infer the frequencies of the transmission maxima as
follows. For a given incidence angle $\theta$, resonances occur at
the intersection of the dispersion relation curve $C$, and the
line $L_{\theta}$, given by $f=(c/2\pi\,\textrm{sin}\theta)
\textrm{Re}[k_x]$. In fact, this is only approximately correct
because the poles lie in the complex plane, and since $L_{\theta}$
is real it cannot intersect a complex $C$. However, whenever
$L_{\theta}$ comes close to $C$, a resonance shows up in the
transmission spectrum. The $\theta=0^{\circ}$ case is exceptional:
as can be observed in Fig.~\ref{bands}, in this case there is an
intersection of $L_{\theta=0^{\circ}}$ and $C$ at
$f_\textrm{odd}=0.783\, c/\Lambda$. But the mode at this frequency
has odd parity, it cannot couple with the even excitation, and
therefore it does not appear in the normal incidence spectrum.
This is verified in Fig.~\ref{oddmode}(a) that renders the
transmittance spectrum for normal and slightly \emph{off}-normal
incidences. The resonant feature at $f_\textrm{odd}$ associated to
the odd mode disappears for normal incidence. The broad maximum
seen in this graph at $f_\textrm{max}$ arises because
$L_{\theta=0^{\circ}}$ comes close to $C$ at this frequency
(symmetry not preventing the coupling in this case). To summarize,
the existence of the surface wave is responsible for \emph{both}
the maximum and the minimum occurring in the transmittance
spectrum for normal incidence (this resonant feature is a typical
Fano resonance profile~\cite{Genet}). Therefore it should not be
concluded that the surface wave is detrimental for the EOT but
rather responsible for it (remind that in the absence of the
dielectric film, the spectrum is absolutely featureless in the
considered range). Let us point out that at $f_\textrm{min}$
[Fig.~\ref{oddmode}(b)] the field in the slits is very close to
zero, and therefore the global field is very similar to the
unperturbed surface wave; this is why $f_\textrm{min}\approx
f_\textrm{unpert}$ (notice, however, that $f_\textrm{min}\neq
f_\textrm{unpert}$). On the other hand, at $f_\textrm{max}$
[Fig.~\ref{oddmode}(c)] the signature of the unperturbed surface
wave can be clearly recognized, but the stronger coupling to
radiation modifies the pattern and consequently shifts the
position of the maximum. The modal field at $f_\textrm{odd}$ can
be inferred from Figs.~\ref{oddmode}(d) and (e).

\begin{figure}[b]
\includegraphics{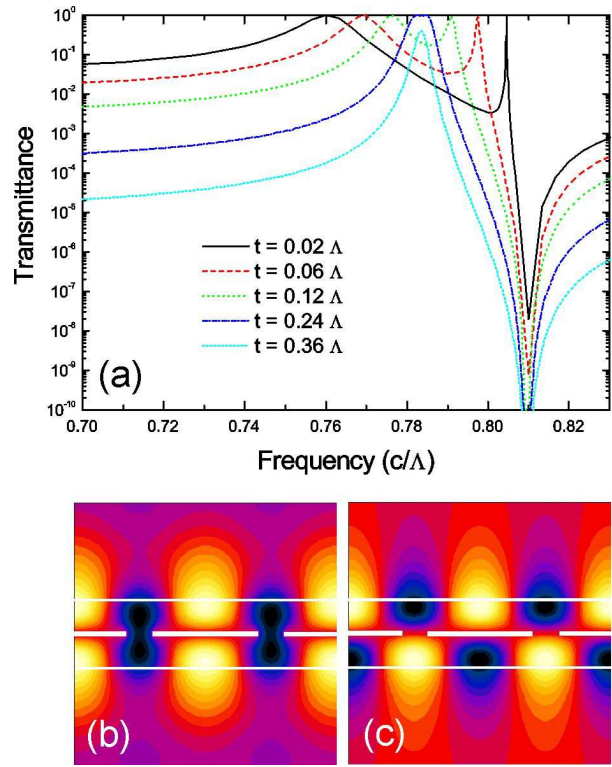}
\caption{\label{symmetric}(color online) (a) Transmittance as a
function of frequency for a \emph{s}-polarized plane wave at
normal incidence. The configuration is symmetric, i.e., there are
two identical films below and above the metallic film
($h=0.25\Lambda$, $\epsilon_\textrm{r}=4$). The slit width is
$w=0.22\Lambda$, and the metal thickness varies between
$t=0.02\Lambda$ and $t=0.36\Lambda$. Panels (b) and (c) display
the electric field for the case $t=0.02\Lambda$ at the left and
right maxima, respectively (the size of the plots is $2\Lambda
\times 2\Lambda$; the color scales are different in the two
panels).}
\end{figure}

At the resonance a high transmittance (but lower than 1) is
obtained. It was already observed in~\cite{Krishnan} that the EOT
can be boosted for symmetric structures. For the configuration
presented here, it is possible to achieve unit transmittance by
considering a symmetric structure, i.e., with identical dielectric
films on the top and the bottom of the metallic film (the slits
are also filled with $\epsilon_\textrm{r}=4$ dielectric, but they
are still subwavelength in the interesting frequency range). Such
a situation is shown in Fig.~\ref{symmetric}. For symmetry
reasons, now four modes are expected, which are even or odd with
respect to both the vertical and the horizontal symmetry planes.
In a similar way to the previously considered case, only two of
them [even with respect to the vertical symmetry plane, see
Fig.~\ref{symmetric}(b) and (c)] show up in the transmittance
spectrum for normal incidence, but now with 100\% transmittance.
When the metal film thickness is increased, the two maxima become
closer, merge, and then the maximum transmittance decreases
exponentially as the metal thickness grows (similar behavior was
found in the 2D hole array case).

\begin{figure}[t]
\includegraphics{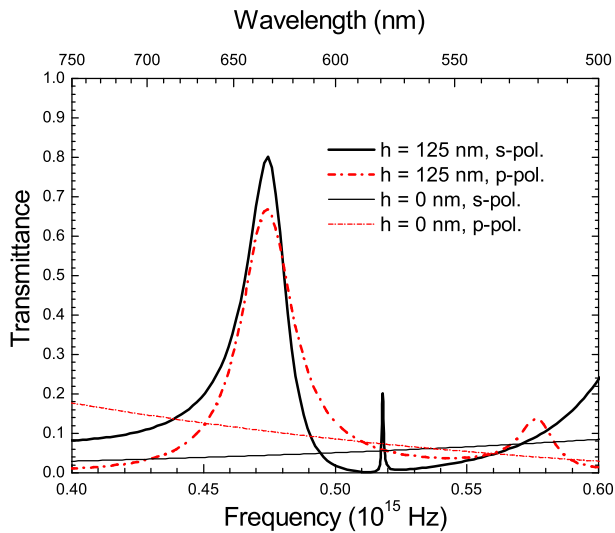}
\caption{\label{realmetal}(color online) Transmittance as a
function of frequency for a normally incident plane wave. The
computation considers a realistic metal in the optical regime, see
main text. Black lines: \emph{s}-polarization. Red lines:
\emph{p}-polarization. The configuration is symmetric, i.e., with
dielectric films at both sides. The dielectric film thickness is
$h=125\,\textrm{nm}$ for the thick lines and $h=0$ for the thin
lines.}
\end{figure}

All our simulations up to this point have considered perfect
metals. The explained mechanism is nevertheless valid for the
optical range, where the perfect metal model breaks down. Let us
therefore conclude with a Drude metal model (appropriate for the
optical regime). The considered parameters are the following:
$\Lambda=400\,\textrm{nm}$, $t=60\,\textrm{nm}$,
$w=110\,\textrm{nm}$, $h=125\,\textrm{nm}$ (in a symmetric
configuration, i.e., dielectric films at both sides), and
$\epsilon_\textrm{r}=4$. The background is vacuum and the metal's
dielectric constant is $\epsilon_\textrm{r,
metal}(f)=1-f^2_\textrm{plasma}/(f^2+if\gamma)$, with
$f_\textrm{plasma}=2.176\times 10^{15}\,s^{-1}$, and
$\gamma=2.418\times 10^{13}\,s^{-1}$, appropriate for gold. As
Fig.~\ref{realmetal} shows (thick black line), the maximum
transmittance (80.5\%) is now lower than in the previous examples.
This is due to the weaker evanescent coupling (the metallic layer
is thicker) and to the absorption losses. The metal supports
surface plasmons and therefore EOT for \emph{p}-polarization is
expected. One would naively expect that this structure is very
sensitive to polarization. However, the parameters can be chosen
so that the \emph{p} and \emph{s}-polarization resonant peaks
occur at the same frequency and with similar transmittances (thick
red line). Thus, a highly polarization-anisotropic structure can
deliver a very polarization-isotropic response.

In conclusion, we have shown that EOT depends uniquely on the
existence of a surface mode and a corrugation allowing the
coupling of the incident wave to the surface mode. EOT for
\emph{s}-polarization (i.e., without plasmons) has been
demonstrated, the nature of the wave and its polarization being
irrelevant for the occurrence of the phenomenon.

Funded by the Spanish MCyT under contracts MAT2002-01534 and
MAT2002-00139, and the EC under projects FP6-2002-IST-1-507879 and
FP6-NMP4-CT-2003-505699.


\end{document}